\documentclass[lettersize,journal]{IEEEtran}
\usepackage{cite}
\usepackage{amsmath,amsfonts}
\usepackage{algorithmic}
\usepackage{array}
\usepackage[caption=false,font=normalsize,labelfont=sf,textfont=sf]{subfig}
\usepackage{textcomp}
\usepackage{multirow}
\usepackage{stfloats}
\usepackage{verbatim}
\usepackage{graphicx}
\usepackage{tipa}
\def\BibTeX{{\rm B\kern-.05em{\sc i\kern-.025em b}\kern-.08em
    T\kern-.1667em\lower.7ex\hbox{E}\kern-.125emX}}
\usepackage{balance}
\begin{document}
\title{ElectrodeNet – A Deep Learning Based Sound Coding Strategy for Cochlear Implants}
\author{Enoch Hsin-Ho Huang, Rong Chao, Yu Tsao, and Chao-Min Wu
\thanks{Manuscript received August xx, 2022; This work was supported by the Ministry of Science and Technology of Taiwan under Grant MOST 108-2221-E-008-066 and 111-2221-E-008-088-MY2. \emph{(Corresponding author: Chao-Min Wu.)}}
\thanks{E. H.-H. Huang is with the Department of Electrical Engineering, National Central University, Taoyuan 320317, Taiwan. He is also with the Research Center for Information Technology Innovation, Academia Sinica, Taipei 115201, Taiwan.}
\thanks{R. Chao is with the Department of Computer Science and Information Engineering, National Cheng Kung University, Tainan 701401, Taiwan. He is also with the Research Center for Information Technology Innovation, Academia Sinica, Taipei 115201, Taiwan.}
\thanks{Y. Tsao is with the Research Center for Information Technology Innovation, Academia Sinica, Taipei 115201, Taiwan, and is a Jointly Appointed Professor with the Department of Electrical Engineering, Chung Yuan Christian University, Taoyuan 320314, Taiwan (e-mail: yu.tsao@citi.sinica.edu.tw).}
\thanks{C.-M. Wu is with the Department of Electrical Engineering, National Central University, Taoyuan 320317, Taiwan (e-mail: wucm@ee.ncu.edu.tw).}}

%\markboth{Journal of xx,~Vol.~xx, No.~x, August~2022}%
%{How to Use the IEEEtran \LaTeX \ Templates}

\maketitle

\begin{abstract}
ElectrodeNet, a deep learning based sound coding strategy for the cochlear implant (CI), is proposed to emulate the advanced combination encoder (ACE) strategy by replacing the conventional envelope detection using various artificial neural networks. The extended ElectrodeNet-CS strategy further incorporates the channel selection (CS). Network models of deep neural network (DNN), convolutional neural network (CNN), and long short-term memory (LSTM) were trained using the Fast Fourier Transformed bins and channel envelopes obtained from the processing of clean speech by the ACE strategy. Objective speech understanding using short-time objective intelligibility (STOI) and normalized covariance metric (NCM) was estimated for ElectrodeNet using CI simulations. Sentence recognition tests for vocoded Mandarin speech were conducted with normal-hearing listeners. DNN, CNN, and LSTM based ElectrodeNets exhibited strong correlations to ACE in objective and subjective scores using mean squared error (MSE), linear correlation coefficient (LCC) and Spearman’s rank correlation coefficient (SRCC). The ElectrodeNet-CS strategy was capable of producing N-of-M compatible electrode patterns using a modified DNN network to embed maxima selection, and to perform in similar or even slightly higher average in STOI and sentence recognition compared to ACE. The methods and findings demonstrated the feasibility and potential of using deep learning in CI coding strategy.
\end{abstract}


\section{Introduction}

\IEEEPARstart{T}{he} cochlear implant (CI) is a revolutionary device that enables people with severe-to-profound hearing loss to restore a partial sense of hearing \cite{CI_review_zeng2008, CI_review_clark2015, CI_review_wilson2015, CI_challenges_TBME_zeng2017} and hence to improve their cognitive functions \cite{CI_cognition_adult_moberly2019, CI_cognition_children_almomani2021}.
With the design of an external CI sound processor and an internal device to stimulate the auditory nerves in meaningful pulse patterns, the lives of hundreds of thousands of CI users have been changed. 
Contributions from multi-disciplinary researchers have enabled the CI today to help not only lip-reading but also speech recognition in quiet conditions \cite{CI_review_zeng2008, CI_review_clark2015, CI_review_wilson2015, CI_challenges_TBME_zeng2017}.
Apart from the present achievements, challenges of electric hearing are still present in understanding speech under interferences, tonal languages, and music \cite{CI_review_zeng2008, CI_review_clark2015, CI_review_wilson2015, CI_challenges_TBME_zeng2017}.
Furthermore, the difficulties in listening to speech through face masks, at increased physical distance, and on the Internet have been getting noticed during the recent COVID-19 pandemic \cite{Covid_children_HL_Taddei2021, Covid_adult_CI_Perez2022}.
Therefore, the continual improvement for the sound processing of the CI is still in great demand.

Deep learning \cite{DL_review_lecun2015,DL_book_goodfellow2016}, a subset of artificial intelligence (AI) and machine learning, has an increasing influence and potential on the innovation of the CI.
In the recent decade, AI based approaches have changed hearing healthcare \cite{AI_hearing_review_lesica2021} and various aspects of the `AI + CI' research \cite{ML_review_crowson2020}, including prognosis estimation \cite{CI_DataMining_prognosis_Guerra_Jimenez2016}, electrode placement \cite{CI_AI_TBME_electrode_gao2016}, robotic surgery \cite{ML_SVM_Surgery_pile2017}, mapping \cite{ML_mapping_meeuws2017}, and sound signal processing \cite{ML_detect_reverb_desmond2013,ML_dereverb_chu2018}.
In addition, the success of deep learning in general audio processing \cite{ML_acoustics_bianco2019, DL_audio_review_purwins2019} and hearing aids \cite{DL_HA_wang2017, Oticon_beck2021} may also change CI sound processing.

In recent years, some CI sound processing approaches have started to shift from traditional to deep learning based solutions.
Distinctive types of network architectures, including deep (feedforward) neural network (DNN) \cite{NNSE_bolner2016, NNSE_goehring2017,citi_ci2017,citi_ci2018}, convolutional neural network (CNN) \cite{CNN_Interspeech_mamun2019, citi_eas_ci2021,citi_visualcues_ci2021}, and recurrent neural network (RNN) \cite{DRNN_nogueira2016, SE_RNN_goehring2019, LSTM_CI_2021}, have been investigated for the CI.
Among the aforementioned studies, deep learning based speech enhancement approaches outperform traditional ones in both simulations and CI recipients \cite{NNSE_bolner2016, NNSE_goehring2017, citi_ci2017, citi_ci2018, CNN_Interspeech_mamun2019}.
Furthermore, multimodal \cite{citi_eas_ci2021, citi_visualcues_ci2021} and music mixing \cite{remix_nogueira2016, remix_nogueira2018, remix_review_nogueira2018} approaches have been demonstrated in performing complex signal processing tasks to improve CI recipients' speech intelligibility and enjoyment of music, respectively.
However, these deep learning based approaches typically modify the front-end processing only.
For the remaining signal processing, particularly for the core sound coding strategy, the efficacy of utilizing deep learning approaches has to be thoroughly investigated.

The CI sound coding strategy plays a key role in representing acoustic signals as recognizable stimulation patterns \cite{strategy_review_wouters2015}.
Continuous interleaved sampling (CIS) \cite{CIS_wilson1991} and advanced combination encoder (ACE) \cite{ACE_original_n_of_m_vandali2000} are two popular coding strategies with reasonably good performance for speech understanding in quiet, but they have already existed for 20-30 years.
More recent commercial coding strategies, including MP3000 \cite{MP3000_PACE_nogueira2005}, fine structure processing (FSP) \cite{FSP_riss2011}, HiRes 120 and Optima \cite{HiRes120_brendel2008, HiRes120_vs_HiRes_firszt2009}, and CRYSTALIS \cite{OticonNeuro_CRYSTALIS_schramm2020}, have been proposed with some advancements.
However, most coding strategies, typically based on traditional signal processing, still need improvements \cite{strategy_review_wouters2015, NCU_CI_TNSRE_huang2021}.

This study aims to identify whether a neural network can perform the signal processing procedures of traditional coding strategies.
In the AI era, the unknown performance and benefits of using deep learning based coding strategies at the expense of computation are of interest for understanding.
Furthermore, the core signal processing techniques of traditional coding strategies, envelope detection (extraction) and channel selection (CS), are not differentiable, and hence unable to be directly combined and optimized with other neural network based front-end and post-processing modules in a joint-training and end-to-end learning manner \cite{single_ch_SE_hansne2010, one_pass_fujimoto2019, input_switch_sato2022}.
Consequently, ElectrodeNet was proposed to demonstrate the concepts of deep learning based CI coding strategy using DNN, CNN, and long short-term memory (LSTM) network models to imitate the envelope detection of the ACE strategy \cite{NCU_CI_ElectrodeNet_huang2019}.
A comprehensive study was carried out for the performance correlation between ElectrodeNet and ACE using objective evaluation with vocoded sentences for the factors of network architecture, dataset language, and noise type.
Subjective listening tests were conducted with normal-hearing Mandarin speakers.

Since the neural network based envelope detection process in ElectrodeNet might not fit well to the conventional CS process, ElectrodeNet-CS, an extended variant for ElectrodeNet containing a CS layer in a modified DNN network, was further proposed.
The ElectrodeNet-CS strategy was trained to optimize the overall CS decision results in attempt to represent speech features in the selected channels.
With the advantage of formulating envelope detection and channel selection in a single network, even slight improvements in speech intelligibility for ElectrodeNet-CS in comparison to ACE and ElectrodeNet would be inspiring.
The current work motivates future research attempting to integrate various front-end and post-processing with ElectrodeNet for different tasks and be optimized in a joint-training and end-to-end learning manner.
We also anticipate that future CI devices will consist of multiple neural network modules, each handling a specific task, such as envelope detection, speech denoising, and acoustic scene classification. 
While current network models appear to require relatively large computational costs, advances in integrated circuits should overcome these limitations \cite{DL_edge_chen2019}.
Therefore, this research can be considered as a pioneering work investigating the feasibility of ElectrodeNet other than its integration with front-end processing units, such as speech denoising and music separation, and as a preparation for the next-generation signal processing of the AI-powered CI system.

The rest of this article is organized as follows.
Section II describes the proposed ElectrodeNet and experimental methods.
Section III presents the objective and subjective evaluation results.
Section IV provides the discussions.
Section V draws the conclusions.

\section{Methods}

\subsection{Strategy Architecture}

This section reviews the fundamentals of the ACE strategy and artificial neural networks, and describes the design of the ElectrodeNet and ElectrodeNet-CS coding strategies.

\subsubsection{ACE}

\begin{figure*}[ht]
\centering{\includegraphics{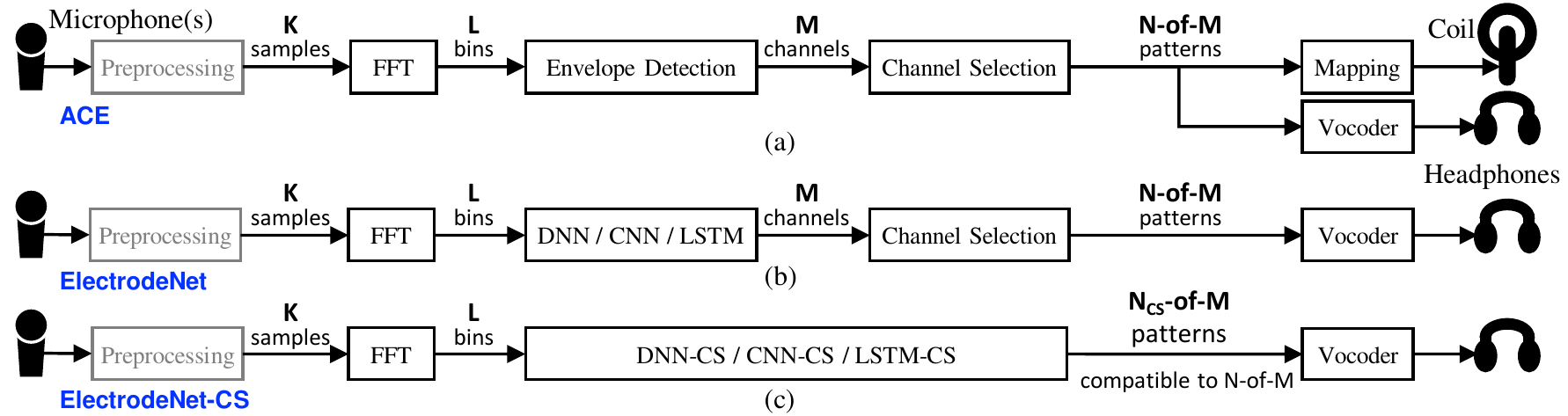}}
\caption{Signal processing for the (a) ACE, (b) ElectrodeNet, and (c) ElectrodeNet-CS coding strategies. The ElectrodeNet strategy uses a neural network, such as a DNN, CNN, or LSTM, to mimic the envelope detection of the ACE strategy. The data formats between the processing stages are denoted as K samples, L bins, M channels, and the N-of-M electrode stimulation patterns. The ElectrodeNet-CS strategy uses a single network containing a custom topk layer to replace both the envelope detection and channel selection stages of ACE and produces the N\textsubscript{CS}-of-M patterns compatible to the N-of-M patterns (N\textsubscript{CS} $\leq$ N). The networks for ElectrodeNet-CS are referred to as DNN-CS, CNN-CS, and LSTM-CS. The preprocessing stages in gray are bypassed in this study.}
\label{fig1_ACE_ENet}
\end{figure*}

ACE \cite{ACE_original_n_of_m_vandali2000}, the most common coding strategy used for Nucleus CI systems, is depicted in Fig.\,\ref{fig1_ACE_ENet} (a).
Sounds collected by microphones are enhanced in the preprocessing stage.
For the Fast Fourier Transform (FFT) filterbank, each short-time frame of generally K = 128 samples is Hann windowed and converted into L = (K/2+1) = 65 bins, where the magnitudes of the complex bins represent the spectral envelopes below 8 kHz.
The envelope detection stage uses the power-sum combination method to merge L bins into M channels \cite{PhDthesis_pitch_swanson2008}, where M = 22 intra-cochlear electrodes for the Nucleus-24 implant with two extra-cochlear ground electrodes.
The CS stage, also known as N-of-M or maxima selection, picks N maxima with the largest amplitudes from M channels.
The mapping stage adjusts stimulation levels based on the loudness growth function (LGF) and personalized electric current units.
Pulse patterns are transmitted from a coil via radio frequencies to the internal implant and may be synthesized as audible sounds for normal-hearing listeners using vocoders \cite{vocoder_shannon1995, vocoder_tone_dorman2002}. 
The ACE strategy defined in Cochlear's Nucleus MATLAB Toolbox (NMT) v4.31 \cite{NMT_swanson2006} was used in this study.

\subsubsection{Artificial Neural Network}

Artificial neural networks, originally simulate the human neural system, have been developed into many distinct architectures over decades.
A network consists of a number of neuron layers, and each neuron combines multiple inputs as a nonlinear output through an activation function.
In the supervised learning procedure, the network is shaped by adjusting its interconnection weights based on labelled data using back-propagation.
The test data is then passed through the learned network to obtain the predicted results in a process called model inference.

DNN, CNN, and LSTM are three commonly used deep learning networks.
A DNN \cite{DNN_BP_werbos1981} or the multilayer perceptron (MLP) \cite{MLP_Hornik1989}, is a feedforward network containing multiple hidden layers or dense layers, where neurons between adjacent layers are fully interconnected.
A CNN typically consists of convolutional layers, pooling layers, and dense layers, and the convolution process is defined by the kernel size, padding, and stride \cite{CNN_lecun1989}.
The LSTM network \cite{LSTM_hochreiter1997} is a refined type of RNN \cite{RNN_rumelhart1986} and designed to use cells and gates to operate as memory in processing sequences with different lengths over time.

\subsubsection{ElectrodeNet}

ElectrodeNet utilizes the artificial neural network in Fig.\,\ref{fig1_ACE_ENet} (b) to imitate the CI electrode output behaviors of the ACE strategy in Fig.\,\ref{fig1_ACE_ENet} (a).
The same FFT filterbank and CS stages were preserved.
Using supervised learning, a DNN, CNN or LSTM model learned to mimic the envelope detection of ACE.
The network input layer was set to L = 65 neurons, which corresponded to the L real-valued bins of FFT magnitudes in ACE.
Paired spectral-temporal matrices in L bins and M channels for clean speech processed by ACE were used to train each neural network.
Electrode stimulation patterns were synthesized as audible sounds using the tone vocoder of NMT and normalized in sound levels.
The deep learning networks were implemented with PyTorch 1.1.0, and the rest of the processing was developed in MATLAB R2020a.
To focus on the performance of coding strategies, the preprocessing stages in Fig.\,\ref{fig1_ACE_ENet} were bypassed in this study.

The neural networks for ElectrodeNet were configured using hyperparameters, including the number of neurons and layers, and some learning settings.
Table \ref{tab1_Network_Architecture} shows the architectures of the DNN, CNN, and LSTM networks investigated.
The DNN model had four dense layers containing 1024, 512, 256, and 22 neurons, respectively.
The CNN model had two layers of 1-D convolution, including 1024 channels and 512 channels with a kernel size of 3 and padding of 2.
The convolutional layers were followed by two dense layers consisting of 256 neurons and 22 neurons, and no pooling layers.
The LSTM network used one 1-D convolution with 1024 channels, one unidirectional LSTM layer with 512 nodes, and two fully connected layers with 256 and 22 neurons, respectively.
For the numbers of parameters shown in Table \ref{tab1_Network_Architecture}, the DNN is the smallest model and the LSTM is the largest model. 
The common activation function was the rectified linear unit (ReLU) \cite{ReLU_maas2013}.
All the three models were trained using the $L_{1}$ norm (mean absolute error, MAE) as the loss function against the ACE envelope detection phase.
The Adam optimizer \cite{Adam_kingma2015} was used with a learning rate of 0.0001, and the number of iterations was 100 epochs.
Causal implementations were used to fit the scenario of real-time signal processing.

\begin{table}
\caption{Network Architecture and Number of Parameters}
\label{tab1_Network_Architecture}
\centering
\begin{tabular}{c|cccc|c}
\hline \hline
Network & DNN & CNN & LSTM & DNN-CS & Size \\ \hline
\multirow{6}{*}{Layer} & input & input & input & input & 65 \\
                       & dense 1 & conv. 1 & conv. 1 & dense 1     & 1024 \\
                       & dense 2 & conv. 2 & lstm 1   & dense 2     & 512 \\
                       & dense 3 & dense 1 & dense 1  & dense 3 & 256 \\
                       & dense 4 & dense 2 & dense 2  & dense 4 & 22 \\
                       & -       & -       & -                         & custom CS & 22 \\ \hline
\# Params & 0.73M & 1.91M & 3.49M & 0.73M & - \\
\hline \hline
\multicolumn{6}{l}{%
  \begin{minipage}[t][0.6cm][t]{8.4cm}%
    \footnotesize Note: The DNN, CNN, and LSTM networks share the same size for different types of network layers.
    The DNN-CS network includes a custom CS layer based on PyTorch's topk function.
  \end{minipage}%
}\\
\end{tabular}
\end{table}

\subsubsection{ElectrodeNet-CS}

\begin{table}
\caption{Percentage for number of channels selected by ElectrodeNet-CS among TMHINT speech frames}
\label{tab2_Percentage_no_of_channels_ElectrodeNet_CS}
\centering
\begin{tabular}{ccccc}
\hline \hline 
Network & Network Maxima & \multicolumn{3}{c}{\# Channels Selected (N\textsubscript{CS}) (\%)} \\ \cline{3-5}
Type & (N\textsubscript{topk}) & $<$ N\textsubscript{topk} & $=$ N\textsubscript{topk} & $>$ N\textsubscript{topk} \\ \hline
\multirow{4}{*}{DNN-CS} & 8, 9, 10, 11 & 0 & 100 & 0 \\
& 12 & 1.98 & 98.02 & 0 \\
& 13 & 1.91 & 98.09 & 0 \\
& 14 & 0.01 & 99.99 & 0 \\
\hline \hline 
\end{tabular}
\end{table}

The ElectrodeNet-CS strategy, a refined variant of ElectrodeNet including the CS stage, is depicted in Fig.\,\ref{fig1_ACE_ENet} (c).
The DNN, CNN, or LSTM network used in ElectrodeNet in Fig.\,\ref{fig1_ACE_ENet} (b) is modified as a DNN-CS, CNN-CS, or LSTM-CS network, which includes an extra CS layer implemented using Pytorch's topk function \cite{PyTorch_topk_web} to select the N\textsubscript{topk} largest channels with positive values.
In contrast to the independent CS process in ElectrodeNet, the CS layer here takes part in the training and optimization process of the neural network weights and the ReLU function was used subsequently to the network model.
Table \ref{tab1_Network_Architecture} shows that the DNN-CS model contains the four dense layers of 1024, 512, 256, and 22 neurons as the DNN model and an extra custom topk layer of 22 neurons.
The number of parameters is equivalent between the DNN model and the DNN-CS model.
The training dataset used for the DNN based ElectrodeNet, paired spectral-temporal matrices in L bins and M channels for clean speech processed by the ACE strategy in Fig.\,\ref{fig1_ACE_ENet} (a), was also used to trained the DNN-CS model.
The training configuration was designed to preserve the most important information in the N\textsubscript{topk} channels selected by the network and to allocate data with smaller envelopes to the remaining M - N\textsubscript{topk} channels using the $L_{1}$ norm computed across the entire M channels.

The ElectrodeNet-CS strategy essentially conformed to the prerequisite N-of-M stimulation patterns of the ACE strategy.
The numbers of channels selected by the DNN-CS network for each speech frame, N\textsubscript{CS}, was variable and mostly equal to the N\textsubscript{topk} setting of the CS layer.
Table \ref{tab2_Percentage_no_of_channels_ElectrodeNet_CS} shows the percentage of N\textsubscript{CS} channels selected among all the Mandarin speech frames used for testing in this study.
For N\textsubscript{topk} = 8, 9, 10, and 11, the N\textsubscript{CS} selected were always equivalent to N\textsubscript{topk}.
For the other DNN-CS networks, the 98.02\%, 98.09\%, and 99.99\% of speech frames were equivalent to the design as N\textsubscript{CS} = N\textsubscript{topk} = 12, 13, and 14, respectively. 
Less than 2\% of the speech frames for the three cases were with fewer channels than design (N\textsubscript{CS} $<$ N\textsubscript{topk}).
Since the number of channels selected were always no greater than the target maxima setting (N\textsubscript{CS} $\leq$ N\textsubscript{topk}), the corresponding N maxima in the ACE strategy, the N\textsubscript{CS}-of-M stimuli generated by the ElectrodeNet-CS strategy were compatible with the conventional N-of-M stimulation patterns.

\subsection{Datasets}

Mandarin and English datasets were both used to investigate a tonal language and a non-tonal language.
The sampling frequency for both datasets was 16 kHz.

\subsubsection{Mandarin dataset}

The Taiwan Mandarin Hearing in Noise Test (TMHINT) sentences \cite{MHINT_wong2007} were used for both objective and subjective evaluations.
The TMHINT material contains 16 phonemically balanced lists, each list has 20 sentences, and each sentence has 10 syllables.
Each syllable corresponds to a Chinese character with one of four primary lexical tones, tone 1 (high and flat), tone 2 (rising), tone 3 (falling then rising), and tone 4 (falling), or a neutral tone.
Sound files recorded by Academia Sinica (AS) and the National Central University (NCU) were used for ElectrodeNet training and testing, respectively.
In training, 320 clean sentences from Lists 9-16 of the TMHINT(AS) material spoken by a male and a female were used.
Specifically, every two lists uttered by one of four male or female speakers were used for training and 160 sentences were for each gender (4 speakers $\times$ 2 lists/speaker $\times$ 20 sentences/list).
In objective evaluation, Lists 1-8 of the TMHINT(NCU) material spoken by one male and one female (16 lists) were mixed with four types of noises, including speech shaped noise (SSN), white noise, street noise and cafeteria babble, at seven signal-to-noise ratio (SNR) levels (-10, -5, 0, 5, 10, 15 dB, and quiet).
For the listening tests, female sentences at four SNR levels (-5, 0, 5 dB, and quiet) with SSN were used.
Sound levels were normalized between the TMHINT(AS) and TMHINT(NCU) sentences.

\subsubsection{English dataset}
An English corpus was used in objective evaluation.
Developed by research teams at Texas Instruments (TI) and the Massachusetts Institute of Technology (MIT), TIMIT is a famous speech dataset used for automatic speech recognition (ASR) \cite{TIMIT_garofolo1993}.
The complete TIMIT dataset consists of 6,300 sentences spoken by 630 different speakers from 8 major dialect regions in the United States.
This study utilized 300 and 80 non-repetitive sentences (150 and 40 speakers / gender) for training and testing, respectively.

\begin{figure*}[ht]
\centering{\includegraphics{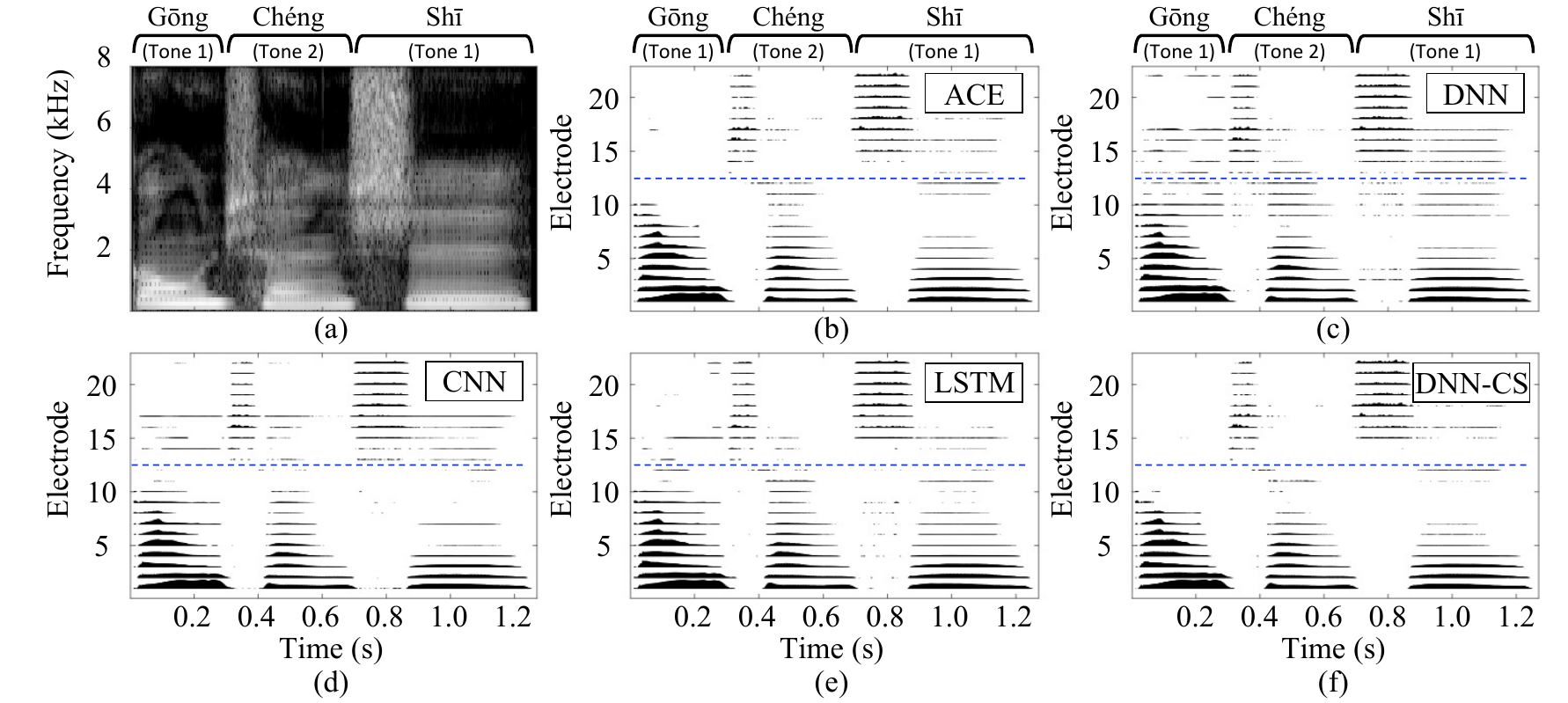}}
\caption{(a) Spectrogram of trisyllabic Mandarin phrase ``Gōng Chéng Shī'' (``Engineer'' in Chinese with tones 1, 2 and 1) uttered by a female speaker. The electrodograms are processed by (b) the ACE strategy and ElectrodeNets based on (c) DNN, (d) CNN, (e) LSTM, and (f) DNN-CS models. The blue dotted lines indicate similar stimuli for the ACE strategy and the DNN-CS based model below Channel 12 (1,937 Hz).}
\label{fig2_Electrodogram}
\end{figure*}

\subsection{Electrode Stimulation Patterns}

Speech signals processed by the ACE and ElectrodeNet strategies with 12 maxima are illustrated in Fig.\,\ref{fig2_Electrodogram}.
The trisyllabic Mandarin phrase ``Gōng Chéng Shī'' (``Engineer'' in Chinese with tones 1, 2 and 1) \cite{Mandarin_trisyllabic_word_nissen2007} has syllabic transitions at approximately 0.3 and 0.7 seconds.
The spectrogram in Fig.\,\ref{fig2_Electrodogram} (a) shows that the consonants of the second and third syllables mainly occur above 2 kHz and the three vowels with strong envelopes exist below 1 kHz.
Similar trends can be observed in the electrodograms for the ACE strategy in Fig.\,\ref{fig2_Electrodogram} (b) and ElectrodeNets based on DNN, CNN, LSTM, and DNN-CS networks in Fig.\,\ref{fig2_Electrodogram} (c), (d), (e), and (f) respectively.
In the first syllable ``Gōng," the `g' consonant (unaspirated velar voiceless stop /k/) is quite short to be clearly identified in all subfigures.
The second and third consonants, `ch' (aspirated retroflex voiceless affricate /t\textrtails\textsuperscript{h}/) and `sh' (retroflex voiceless fricative /\textrtails/), are in lengths of approximately 100 ms (0.3-0.4s in the time axis) and 200 ms (0.7-0.9s in the time axis), respectively.
At the lower frequencies from Channel 1 (250 Hz) to Channel 8 (1,125 Hz), the pulses with higher amplitudes represent vowels carrying strong energy in comparison to relatively weak consonants.
Some differences between ACE and ElectrodeNets, particularly the DNN case in Fig.\,\ref{fig2_Electrodogram} (c), can be observed between Channel 9 (1,250 Hz) and Channel 17 (3,812 Hz) for the three vowels, but these pulses in the mid-to-high spectrum should only have small effect on vowels dominated by low frequencies.

The electrodogram for the ElectrodeNet-CS strategy based on the DNN-CS model is generally similar to that for ACE.
In Fig.\,\ref{fig2_Electrodogram}, the electrode stimulation patterns under the blue dotted lines are similar between the ACE strategy and the ElectrodeNet-CS strategy compared to the other three ElectrodeNets without the CS network layer.
Specifically, the stimuli for ACE and ElectrodeNet-CS exhibit comparable patterns below Channel 12 (1,937 Hz), the low-to-mid frequency range important for speech recognition.

\subsection{Objective Evaluation}

This section explains the objective predictors, experimental factors, and correlation computations for objective evaluation.

\subsubsection{Predictors}

Two objective predictors, short-time objective intelligibility (STOI) \cite{stoi_taal2011} and normalized covariance metric (NCM) \cite{ncm_holube1996, ncm_goldsworthy2004} were used to evaluate ElectrodeNet's performance in speech intelligibility.
Objective predictors have been widely used in estimating speech intelligibility, quality, and output SNR for assistive hearing devices \cite{ncm_evaluate_chen2011, SRMR_objective_review_loizou_santos2013, objective_evaluation_review_falk2015, OSNR_watkins2018}.
Among these predictors, STOI and NCM are two intrusive predictors for speech intelligibility using a reference speech and a processed version of the speech.
STOI has a high correlation with speech intelligibility and has been used in CI studies on speech enhancement and coding strategies \cite{objective_evaluation_review_falk2015, Sparse_NMF_hu2015, citi_ci2017, MOCstrategy_lopez_poveda2018, MOCstrategy_lopez_poveda2020, VG_langner2020, TFMask_mourao2020, citi_visualcues_ci2021, citi_eas_ci2021,NCU_CI_TNSRE_huang2021}.
NCM is another frequently used indicator for evaluating speech comprehension in CI sound processing \cite{objective_evaluation_review_falk2015, Sparse_NMF_hu2015, citi_ci2017, TFMask_mourao2020, citi_visualcues_ci2021}, including for vocoded Mandarin speech \cite{ncm_evaluate_chen2011}.
In this study, STOI or NCM scores were computed between the clean original speech and the corresponding processed and vocoded speech.

\subsubsection{Experimental Factors}

The ElectrodeNet strategy was objectively evaluated for experimental factors of network architecture, dataset language, and noise type.

\begin{itemize}
\item{\textbf{Network Architecture}}

The DNN, CNN, and LSTM based ElectrodeNets were investigated to clarify the efficacy of the deep, convolutional, and recurrent network architectures.

\item{\textbf{Dataset Language}}

The DNN based ElectrodeNet was trained and tested using the TMHINT dataset or the TIMIT dataset to evaluate the objective performance with different languages.

\item{\textbf{Noise Type}}

The DNN based ElectrodeNet was evaluated for speech understanding with four types of noises (SSN, white noise, street noise \cite{SE_book_loizou2013} and cafeteria babble \cite{SE_book_loizou2013}) at seven SNR levels (-10, -5, 0, 5, 10, 15 dB and quiet).

\end{itemize}

The three factors were examined for their effects on the correlations of objective speech intelligibility between ElectrodeNet and ACE.

\subsubsection{Correlations in Objective Speech Intelligibility}

The relationships between ElectrodeNet and the ACE strategy in speech intelligibility were evaluated using mean squared error (MSE), linear correlation coefficient (LCC) \cite{LCC_pearson1920}, and Spearman’s rank correlation coefficient (SRCC) \cite{SRCC_spearman1904}.
The MSE measured the average of the squares of the differences in speech intelligibility scores between ACE and ElectrodeNet.
The closer to 0 the MSE scores, the smaller the errors between the two strategies.

The LCC between ACE and ElectrodeNet is computed as

\begin{equation}
\textrm{LCC}=\rho=\frac{\textrm{cov}(SI_{ACE},SI_{ElectrodeNet})}{\sigma_{SI_{ACE}}\sigma_{SI_{ElectrodeNet}}}
\label{Equation: LCC}\end{equation}

where $SI$ is the speech intelligibility for a coding strategy, cov$(X,Y)$ is the covariance of $X$ and $Y$, and $\sigma$ is the standard deviation.
Therefore, the LCC is the covariance of speech intelligibility for ACE and ElectrodeNet divided by the product of standard deviations of speech intelligibility for the two strategies.
The SRCC is computed as

\begin{equation}
\textrm{SRCC}=\rho_s=\frac{\textrm{cov}(Rank({\scriptstyle ACE}),Rank({\scriptstyle ElectrodeNet}))}{\sigma_{Rank(ACE)}\sigma_{Rank(ElectrodeNet)}}
\label{Equation: SRCC}\end{equation}

where $Rank(\cdot)$ is the resulting rank in speech intelligibility for one coding strategy.
The ranks for $n$ sentences are generally distinct integers $1, 2, ..., n$ with respect to the ascending order of speech intelligibility, and average ranks are used for repeated intelligibility scores.
In both equations (\ref{Equation: LCC}) and (\ref{Equation: SRCC}), the order of the two coding strategies does not change the values of LCC and SRCC.
Typically, a correlation coefficient greater than 0.7 or 0.8 indicates a strong correlation \cite{Correlation_Strong_0p8_zou2003, Correlation_Strong_0p7_akoglu2018}.
The closer the LCC and SRCC scores to 1, the higher the correlations in speech intelligibility between the two strategies.
Consequently, the desired results for correlations between ElectrodeNet and ACE are small MSEs close to 0 and large LCCs and SRCCs close to 1.

\subsubsection{ElectrodeNet-CS}

The ElectrodeNet-CS strategy based on the DNN-CS model was evaluated using objective predictors for maxima settings N = 8 and N = 12.
The STOI and NCM scores were measured for ACE, the DNN, CNN, and LSTM based ElectrodeNets, and the DNN-CS model based ElectrodeNet-CS strategy.
Statistical analyses were conducted for average list scores using IBM SPSS Statistics 21 (released 2012) with a significance level of $\alpha$ = 0.05.

\subsection{Subjective Evaluation}

Two listening experiments were conducted using CI simulations.
The first experiment focused on the speech intelligibility of the DNN and CNN based ElectrodeNets, and the second one evaluated the ElectrodeNet-CS strategy.
The procedures were approved by an institutional review board (IRB).

\subsubsection{Listening Experiment 1}
Listening tests were performed with two groups of normal-hearing listeners to investigate the correlations between ACE and two ElectrodeNet strategies.
Forty native Mandarin speakers (age 20-29) were randomly divided into two test groups to evaluate the performance of the DNN or CNN based ElectrodeNet compared to the ACE strategy.
In the DNN group, 14 male and 6 female subjects were with an average age of 22.3 years, while 15 male and 5 female subjects with an average age of 22.6 were assigned to the CNN group.
All subjects had hearing thresholds no more than 25 dB HL at octave frequencies between 0.125 and 8 kHz in pure tone audiometry (PTA).
The sentence recognition test for each subject took about 45-60 minutes.

Speech intelligibility was measured for the ACE strategy and an ElectrodeNet strategy based on the DNN/CNN group assignment.
Each subject listened to the simulated CI sound simultaneously played to both ears via Telephonics TDH-39P headphones in a sound isolation room.
The experiment was carried out with NCU-CI, a MATLAB-based CI experiment platform with a graphical user interface in traditional Chinese\cite{NCU_CI_wu2009, NCU_CI_ElectrodeNet_huang2019, NCU_CI_TNSRE_huang2021}.
In the practice session, each subject was instructed on the experimental procedure by being familiarized with sentences from List 16 of the TMHINT material presented at a fixed and comfortable sound level.
Vocoded sentences processed by ACE and ElectrodeNet with maxima N = 12 were presented at four SNR levels in the order of -5, 0, 5 dB and quiet.
At each SNR level, the order of ACE and ElectrodeNet was chosen randomly.
For the eight test conditions (2 strategies $\times$ 4 SNR levels), one of TMHINT Lists 1-8 was selected at random without replacement, and the 20 sentences for each condition were also presented in a random order. 
Each sentence was played only once and the subjects were encouraged to type the Chinese characters to the best of their ability.
The speech intelligibility for each list was calculated as the average percentage of correct results for 200 syllables (10 syllables/sentence × 20 sentences/list).
Scores were given for each correct syllable, including a homophone, a different Chinese character with the correct pronunciation.

\subsubsection{Listening Experiment 2}
Sentence recognition tests were conducted with twenty normal-hearing listeners to investigate the DNN based ElectrodeNet-CS strategy with two maxima configurations.
Ten male and ten female native Mandarin speakers (age range 20-25 years, mean age of 21.7 years) participated the experiment.
Four coding strategy conditions (ACE with N = 8 and 12, and ElectrodeNet-CS with N = 8 and 12) were tested randomly for each SNR level.
TMHINT List 1-8 were randomly tested for both ElectrodeNet-CS strategies.
Owing to the limited sentences for listening tests, Lists 9-16, already used for training the two ElectrodeNet-CS strategies, were tested for the two ACE conditions.
The rest procedure was similar to that used in Experiment 1.
Each subject spent about 60-90 minutes in this experiment.

\section{Results}

\subsection{Objective Evaluation Results}

This section presents the correlations between ElectrodeNet and ACE for the factors of network architecture, language, and noise type, and compares the two coding strategies with the ElectrodeNet-CS strategy in objective speech intelligibility.

\begin{figure}[!t]
\centerline{\includegraphics[width=\columnwidth]{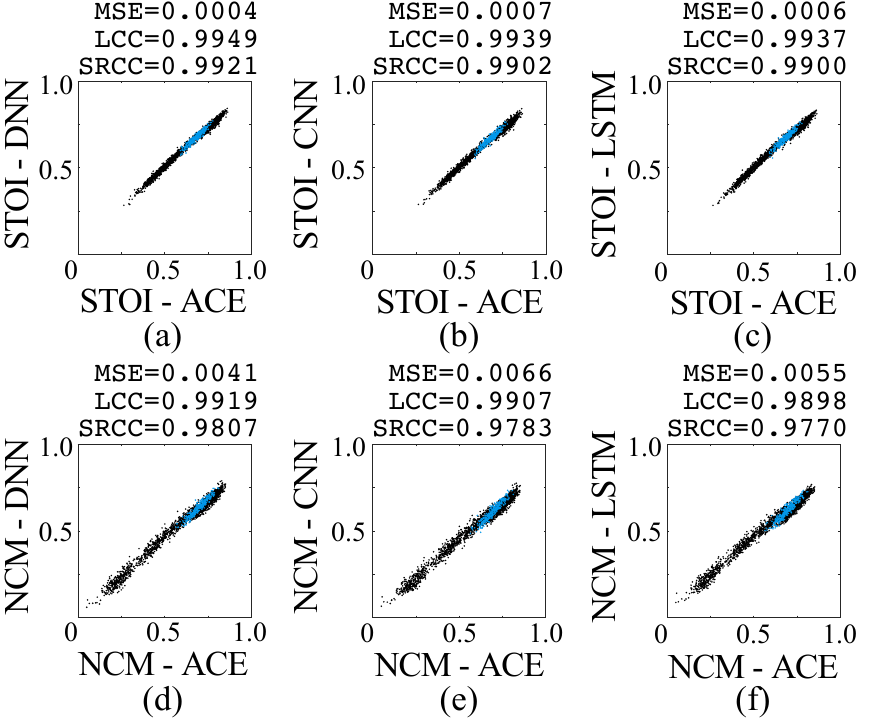}}
\caption{Scatter plots of STOI scores for (a) DNN, (b) CNN, and (c) LSTM networks and NCM scores for (d) DNN, (e) CNN, and (f) LSTM networks compared to ACE under SSN. The blue dots indicate scores at 5 dB SNR.}
\label{fig3_ScatterSTOI_NCM}
\end{figure}

\subsubsection{Network architecture}

The scatter plots in Fig.\,\ref{fig3_ScatterSTOI_NCM} illustrate the paired objective scores of ACE and ElectrodeNet.
In Fig.\,\ref{fig3_ScatterSTOI_NCM} (a)-(c), each dot on a rectangular coordinate system denotes a pair of STOI scores for ACE and ElectrodeNet at $(STOI_{ACE}, STOI_{ElectrodeNet})$ for one vocoded sentence.
Likewise, each dot in Fig.\,\ref{fig3_ScatterSTOI_NCM} (d)-(f) denotes the NCM results estimated for one sentence at $(NCM_{ACE}, NCM_{ElectrodeNet})$.
Similar trends can be noted among the DNN, CNN, and LSTM based ElectrodeNets in both STOI and NCM scores.
To clarify that the correlations are driven by the characteristics of neural networks instead of the effects of noise, the blue dots denoting the 5 dB condition indicate that the relationships are approximately linear for a single SNR level.
For more informative explanations, the scatter plots and correlation scores at each SNR level are included in Appendix.
At low SNR levels, the dots representing the STOI scores are distributed farther from the origin coordinates compared to the NCM results, which suggest different properties for the two objective predictors.

The MSE, LCC, and SRCC scores in Fig.\,\ref{fig3_ScatterSTOI_NCM} reveal evidence of strong correlations between ACE and the DNN, CNN, and LSTM based ElectrodeNets.
All the MSE results based on the STOI predictor were close to 0 in a range of 0.0004–0.0007.
The LCC and SRCC scores were close to the ideal result of 1 in ranges of 0.9937–0.9949 and 0.9900–0.9921, respectively.
For the NCM results, the MSE scores were in a range of 0.0041–0.0066, and LCC and SRCC results were within ranges of 0.9898–0.9919 and 0.9770–0.9807, respectively.
For DNN, the MSEs (0.0004 with STOI and 0.0041 with NCM) were slightly smaller than those for CNN and LSTM (0.0006-0.0007 with STOI and 0.0055-0.0066 with NCM), and the LCC (0.9949 with STOI and 0.9919 with NCM) and SRCC (0.9921 with STOI and 0.9807 with NCM) were slightly greater than those for CNN and LSTM.
Although a slightly stronger relationship with ACE was observed for the DNN based ElectrodeNet compared to the CNN and LSTM based ElectrodeNets, the differences between the scores were quite small.
Consequently, strong relationships existed between the ACE strategy and each of the DNN, CNN, and LSTM based ElectrodeNets.
Further investigations into the three types of ElectrodeNets and the ElectrodeNet-CS strategy using analysis of variance (ANOVA) are discussed in a later section.

\begin{table}[]
\caption{Objective evaluation results for Mandarin and English speech materials using DNNs with SSN}
\label{tab3_TrainingSentenceCorr}
\centering
\begin{tabular}{cccccc}
\hline \hline
Predictor & Language & Speech Material & MSE & LCC & SRCC\\
\hline
\multirow{2}{*}{STOI} & Mandarin & TMHINT & 0.0004 & 0.9949 & 0.9921 \\
 & English & TIMIT & 0.0004 & 0.9951 & 0.9930 \\
\hline
\multirow{2}{*}{NCM} & Mandarin & TMHINT & 0.0041 & 0.9919 & 0.9807 \\
 & English & TIMIT & 0.0031 & 0.9921 & 0.9843 \\
\hline \hline
\end{tabular}
\end{table}

\subsubsection{Language}

Strong relationships between the ACE and ElectrodeNet strategies for Mandarin and English speech are shown in Table \ref{tab3_TrainingSentenceCorr}.
For both datasets, the errors between ACE and ElectrodeNet were small and the correlations between the strategies were reasonably high.
All the MSEs were close to zero (0.0004 for STOI with both languages, and 0.0031-0.0041 for NCM).
The LCCs were greater than 0.99 for all conditions, and the SRCCs were above 0.99 for STOI and 0.98 for NCM.
Therefore, ElectrodeNet appeared to have a high correlation to ACE with both English and Mandarin.

\begin{table}[]
\caption{Objective evaluation results for the DNN model under SSN, white noise, street noise, and cafeteria babble.}
\label{tab4_NoiseTypeCorr}
\centering
\begin{tabular}{ccccc}
\hline \hline 
Predictor & Noise Type & MSE & LCC & SRCC\\ \hline
\multirow{4}{*}{STOI} & SSN & 0.0004 & 0.9949 & 0.9921 \\
& white noise & 0.0006 & 0.9878 & 0.9870 \\
& street noise & 0.0009 & 0.9899 & 0.9837 \\
& cafeteria babble & 0.0005 & 0.9932 & 0.9887 \\ \hline
\multirow{4}{*}{NCM} & SSN & 0.0041 & 0.9919 & 0.9807 \\
& white noise & 0.0035 & 0.9858 & 0.9834 \\
& street noise & 0.0071 & 0.9693 & 0.9609 \\
& cafeteria babble & 0.0060 & 0.9878 & 0.9773 \\
\hline \hline
\end{tabular}
\end{table}

\subsubsection{Noise type}

The DNN based ElectrodeNet exhibited strong correlations with the ACE strategy in STOI and NCM scores under four types of noises as listed in Table \ref{tab4_NoiseTypeCorr}.
The MSE, LCC, and SRCC scores were quite close between SSN, white noise, street noise, and cafeteria babble.
For STOI, the MSEs were in a range of 0.0004–0.0009, and all the LCCs and SRCCs were greater than 0.98.
As to NCM, the MSEs were in a range of 0.0035–0.0071.
The strongest correlations mostly occurred with SNN.
The lowest correlation coefficients occurred for street noise estimated by NCM, where both LCC and SRCC were greater than 0.96.
Hence, ElectrodeNet and ACE were highly correlated in speech intelligibility under all the four types of noises.

\subsubsection{The ElectrodeNet-CS Strategy}

\begin{figure}[!t]
\centerline{\includegraphics[width=\columnwidth]{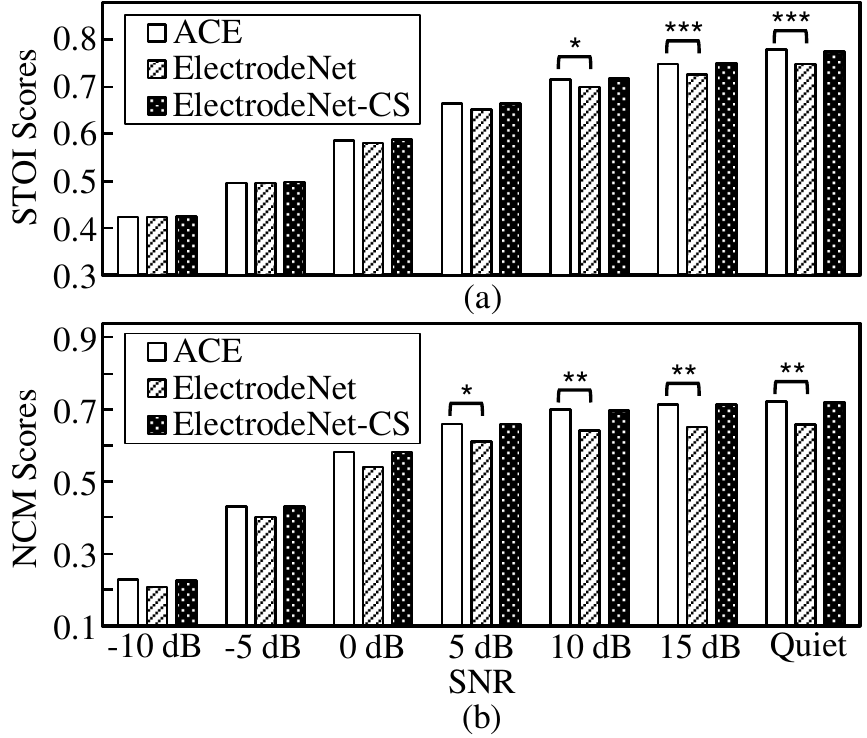}}
\caption{Average (a) STOI scores and (b) NCM scores for vocoded TMHINT sentences processed by the ACE, ElectrodeNet and ElectrodeNet-CS strategies with eight maxima selected (N = 8). Asterisks indicate significant difference to the ACE strategy with \text{*}\emph{p} $\leq$ 0.05, \text{**}\emph{p} $\leq$ 0.01, and \text{***}\emph{p} $\leq$ 0.001.}
\label{fig4_STOI_NCM_ElectrodeNet_CS}
\end{figure}

The ElectrodeNet-CS strategy was compared with the ACE strategy and the DNN, CNN, LSTM, and DNN-CS network based ElectrodeNets for objective speech intelligibility scores.
Fig.\,\ref{fig4_STOI_NCM_ElectrodeNet_CS} shows the average STOI and NCM scores for TMHINT sentences processed by the ACE strategy, the DNN based ElectrodeNet, and ElectrodeNet-CS using eight maxima (N = 8).
The average scores for the CNN and LSTM networks not shown here were slightly lower than those for the other three strategies.
A two-way ANOVA was conducted for 560 average list scores (8 lists $\times$ 2 speakers $\times$ 5 strategies $\times$ 7 SNR levels) with coding strategy and SNR level as factors, and the quiet condition was considered as one of the seven SNR levels.
Significant effects were observed with both coding strategy and SNR level (\emph{p} $<$ 0.001) for both STOI and NCM results.
Post-hoc analysis at each SNR level indicated that the DNN based ElectrodeNet and ACE were distinguishable at SNRs above 10 dB for STOI and above 5 dB for NCM (\emph{p} $<$ 0.05).
The results with the ElectrodeNets based on the CNN and LSTM networks were similar at SNR levels above 5 dB and 10 dB for STOI scores and -5 dB and 0 dB for NCM scores, respectively.
On the other hand, the ElectrodeNet-CS strategy was similar to ACE with both STOI and NCM at all SNR levels in Fig.\,\ref{fig4_STOI_NCM_ElectrodeNet_CS} (\emph{p} $>$ 0.05), and the correlations were extremely high (MSEs $\leq$ 0.00002, LCCs $\geq$ 0.9996, SRCC $\geq$ 0.9995) as listed in Table \ref{tab5_EC_Corr}.
Surprisingly, the average STOI scores for ElectrodeNet-CS were slightly higher than the results for ACE at most SNR levels except for the quiet condition, and the differences in scores were in a small range of 0.0006-0.0021.
Similar results were observed for the N = 12 condition, where the mean STOI scores for ElectrodeNet-CS were slightly higher than those for ACE across all SNR levels in a range of 0.0013-0.0021, except for a slightly lower score by 0.0015 in the quiet condition.
For the average NCM scores, the differences between the two strategies were below 0.0024 for both N = 8 and N = 12.

\begin{table}[]
\caption{Objective evaluation results for the DNN-CS model}
\label{tab5_EC_Corr}
\centering
\begin{tabular}{cccccc}
\hline \hline
Predictor & Network Maxima & MSE & LCC & SRCC\\
\hline
\multirow{2}{*}{STOI} & N\textsubscript{topk} = 8 & 0.00001 & 0.9996 & 0.9995 \\
& N\textsubscript{topk} = 12 & 0.00001 & 0.9998 & 0.9997 \\
\hline
\multirow{2}{*}{NCM} & N\textsubscript{topk} = 8 & 0.00002 & 0.9996 & 0.9995 \\
 & N\textsubscript{topk} = 12 & 0.00001 & 0.9998 & 0.9996 \\
\hline \hline
\end{tabular}
\end{table}

\subsection{Listening Test Results}

\subsubsection{Experiment 1 (ElectrodeNet)}

\begin{figure}[!t]
\centerline{\includegraphics[width=\columnwidth]{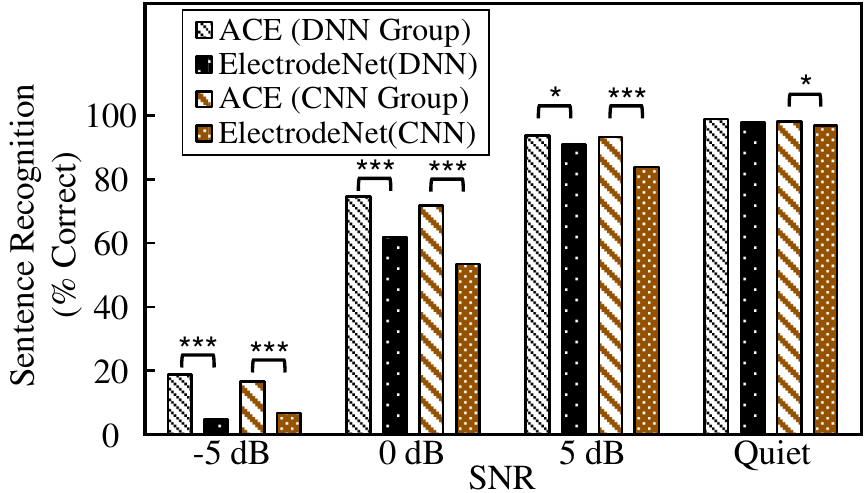}}
\caption{Average Mandarin sentence recognition scores of two groups of 20 NH subjects for the ACE and ElectrodeNet strategies at four SNR levels (N = 12). The DNN group participated listening tests for the ACE strategy and the DNN based ElectrodeNet (in monochrome), and the CNN group for ACE and the CNN based ElectrodeNet (in brown). Asterisks indicate significant differences with \text{*}\emph{p} $\leq$ 0.05 and \text{***}\emph{p} $\leq$ 0.001. No statistical difference was observed between ACE and the DNN based ElectrodeNet in the quiet condition.}
\label{fig5_Subjective}
\end{figure}

The listening test results for the DNN and CNN based ElectrodeNets in comparison to the ACE strategy are illustrated in Fig.\,\ref{fig5_Subjective}.
The average percent correct scores for the DNN group and the CNN group are denoted in monochrome and blue, respectively.
Two-way repeated-measures (RM) ANOVA tests were conducted for the two groups of subjects.
Significant effects were noted for both factors of SNR and strategy (\emph{p} $<$ 0.001).
Paired-samples T-tests indicated that the DNN based ElectrodeNet performed similarly to the ACE strategy in the quiet condition (\emph{p} $>$ 0.05), while the results for the other conditions showed slightly lower speech intelligibility for the DNN and CNN based ElectrodeNets compared to ACE (\emph{p} $<$ 0.05).
At less noisy SNR levels, the differences to the ACE strategy were 3.0\% (5 dB) and 1.0\% (quiet) for the DNN based ElectrodeNet, and 9.5\% (5 dB) and 1.3\% (quiet) for the CNN counterpart.
At -5 dB and 0 dB, the ACE strategy yielded speech intelligibility scores 10.0-18.3\% higher than those for ElectrodeNets.
It appears that only the DNN based ElectrodeNet modeled the ACE strategy well in the quiet condition.

\subsubsection{Experiment 2 (ElectrodeNet-CS)}

\begin{figure}[!t]
\centerline{\includegraphics[width=\columnwidth]{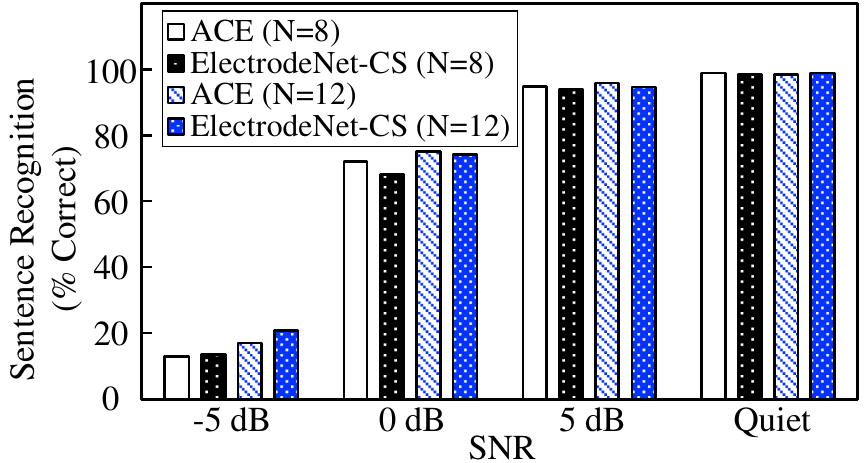}}
\caption{Average Mandarin sentence recognition scores of 20 NH subjects for the ACE and ElectrodeNet-CS strategies with N = 8 (in monochrome) and N = 12 (in blue) at four SNR levels. No significant difference was observed between the two strategies for each N and SNR level (\emph{p} $>$ 0.05).}
\label{fig6_Subjective2_score}
\end{figure}

The sentence recognition results for ACE and ElectrodeNet-CS with 8 and 12 maxima at four SNR levels are shown in Fig.\,\ref{fig6_Subjective2_score}.
For each maxima N, the two-way RM ANOVA showed a significant effect with SNR (\emph{p} $<$ 0.001), but no significant effect with coding strategy (\emph{p} $>$ 0.05).
Paired-samples T-tests for each SNR level and maxima N indicated no significant difference between the two strategies in all comparisons (\emph{p} $>$ 0.05).
At -5 dB SNR, the average percent correct scores of the ElectrodeNet-CS strategy, 13.6\% for N = 8 and 20.8\% for N = 12, were slightly higher than those of the ACE strategy, 13.0\% and 17.0\%, respectively.
For the 0 dB SNR, the mean recognition scores for ACE and ElectrodeNet-CS were 72.1\% and 68.2\% for N = 8, and 75.2\% and 74.2\% for N = 12, respectively.
For the 5 dB SNR and the quiet condition, the average results for ACE and ElectrodeNet-CS were similar for both N maxima, and the differences were within 1.47\%.
The average score for ElectrodeNet-CS with N = 12 in the quiet condition was 99.1\% and slightly higher than the 98.5\% score for ACE.
The experimental and statistical analysis results implied that the ElectrodeNet-CS strategy performed similarly to ACE with Mandarin speech.

\section{Discussion}

\subsection{ElectrodeNet}

The strong correlations between ElectrodeNet and the ACE strategy in STOI and NCM scores suggested that a deep learning based coding strategy is capable of producing stimuli with speech intelligibility closely related to that for ACE.
These close relationships held for the three experimental factors of network architecture, dataset language, and noise type.
By inspecting the architecture of neural networks, the L-to-M envelope detection of ACE can be considered as a special case of a fully-connected network with fixed weights, no hidden layers, no biases, and bypassed activation functions.
In other words, the dense layers in the DNN, CNN, and LSTM models can be regarded as generalized forms of ACE's envelope detection using nonlinear operations.
Consequently, the DNN based ElectrodeNet exhibited a slightly stronger relationship with ACE, possibly due to its closer network architecture compared to CNN and LSTM structures.
In this study, neural networks were trained by a single frame at each time.
If more temporal and spectral information was referenced during training, other networks with more advanced architecture, such as CNN and LSTM, may take advantage of their capabilities in model framework.
In terms of the dataset language factor, the DNN based ElectrodeNet performed fairly well in representing Mandarin and English speech as analogous forms of ACE vocoded speech with TMHINT and TIMIT sentences.
For the factor of noise type, the correlations between the two strategies in STOI and NCM scores were generally stronger for SSN due to its closer resemblance to the training speech materials compared to white noise, street noise, and cafeteria babble.

Although the STOI and NCM scores between ElectrodeNet and ACE showed `strong correlations,' they were found to be `significantly different' in ANOVA analysis.
The results of analyzing identical data using the two different types of statistical measurements appeared to be different, but these discrepancies may still be explained.
The MSE, LCC, and SRCC scores measured the strength of relationship between the two strategies in STOI and NCM scores.
Since the neural networks mimicked the envelope detection function, the outputs of ElectrodeNet were dependent to those of ACE with some relationships, which were quantified as strong correlations in this study.
ANOVA, however, compares the means of different groups of STOI and NCM scores processed by different coding strategies.
Based on supervised learning, the DNN network of ElectrodeNet learned the process of the L-to-M envelope detection, but the network output envelopes may not always closely follow the those of the ACE strategy.
Using network output as envelopes could impact the subsequent N-of-M selection, potentially leading to reduced intelligibility.
For example, the electrodogram in Fig.\,\ref{fig2_Electrodogram} (c) showed that the DNN based ElectrodeNet generated a few undesired high-frequency pulses as interferences and hence may have negative effects on the performance.
Therefore, ElectrodeNet-CS was proposed to investigate whether including a CS layer would improve ElectrodeNet to perform more closely to the ACE strategy.

\subsection{ElectrodeNet-CS}

Both the objective and subjective evaluations showed that the ElectrodeNet-CS strategy had similar or even slightly higher average scores than ACE under certain conditions.
The superior results of the DNN-CS model for ElectrodeNet-CS compared to the DNN based ElectrodeNet were consistent with the electrodograms in Fig.\,\ref{fig2_Electrodogram}.
Compared to the conventional N-of-M process of ElectrodeNet, the custom topk layer of the ElectrodeNet-CS strategy, took part in network training and resulted in a more similar modeling of ACE's channel envelopes, particularly the stimuli important for speech recognition below approximately 2 kHz (Channel 12).

In this study, one aspect that required clarification was the use of a different training target when comparing the signal processing architectures of ACE and ElectrodeNet-CS in Fig.\,\ref{fig1_ACE_ENet}.
For a match to the CS output of the ACE strategy, the N-of-M patterns were adopted as the training target of the DNN-CS model.
Table \ref{tab6_CS_Target} shows that the DNN-CS network trained using the N-of-M selection output of the ACE strategy (N = 8) was with lower STOI scores than using the entire spectrum of M-channel envelopes.
With the N-of-M target, the neural network was trained to prioritize N channels with positive values, but the remaining M - N channels were trained to converge toward zeros.
When a network learned to produce zeros in some channels, the resulting stimuli became sparser and the corresponding STOI scores decreased.
The use of the entire M-channel target for training, however, determined the CS output based on full spectral contents instead of just the N maximal envelopes.
Therefore in this study, the M-channel data was used to train the DNN-CS model for its slightly higher performance compared to the N-of-M patterns.

Another aspect to be clarified was the effectiveness of using PyTorch's topk function directly in the CS layer, referred to as vanilla topk (VT), as opposed to a customized topk.
The network with the VT function, DNN-CS\textsubscript{VT}, produced slightly higher STOI scores, 0.4297 (-10 dB), 0.5899 (0 dB), 0.7213 (10 dB), and 0.7837 (quiet), compared to ACE and the DNN-CS model as listed in Table \ref{tab6_CS_Target}.
However, the NCM scores for DNN-CS\textsubscript{VT} were lower than those for ACE scores by a range of 0.02-0.03, compared to a smaller range of 0.001-0.0024 for the DNN-CS model.
Furthermore, the DNN-CS\textsubscript{VT} strategy switched off some channels while minimizing the loss.
For example, the DNN-CS\textsubscript{VT} model with N\textsubscript{topk} = 12 bypassed Channels 18 (4,375 Hz), 19 (5,000 Hz), and 21 (6,500 Hz), while the model with N\textsubscript{topk} = 8 even bypassed six channels.
Although disabling high-frequency channels may increase STOI scores due to the greater relevance of low-to-mid frequency components to speech perception, the seemingly random deactivation of electrodes was problematic.
In the case of training the DNN-CS\textsubscript{VT} network with the N-of-M target, sparser output resulted in lower STOI scores compared to all other conditions shown in Table \ref{tab6_CS_Target}, suggesting a lack of robustness in the network's structure.
Owing to the aforementioned limitations, a custom topk layer was used in ElectrodeNet-CS, while the findings with the DNN-CS\textsubscript{VT} network presented herein may still offer some insights into related CI research.

\begin{table}
\caption{Average STOI scores for the DNN-CS and DNN-CS\textsubscript{VT} models\\ with different training targets (N = 8)}
\label{tab6_CS_Target}
\centering
\begin{tabular}{ccc|cc|cc}
\hline \hline
SNR & \multirow{2}{*}{ACE} & \multirow{2}{*}{DNN} & \multicolumn{2}{c |}{DNN-CS} & \multicolumn{2}{c}{DNN-CS\textsubscript{VT}} \\ \cline{4-7}
(dB) & & & Target: M & N-of-M & M & N-of-M \\ \hline
-10 & 0.4229 & 0.4234 & 0.4245 & 0.4225 & 0.4297 & 0.4166 \\
0 & 0.5868 & 0.5810 & 0.5884 & 0.5823 & 0.5899 & 0.5614 \\
10 & 0.7163 & 0.7000 & 0.7173 & 0.7011 & 0.7213 & 0.6756 \\
Quiet & 0.7786 & 0.7495 & 0.7766 & 0.7505 & 0.7837 & 0.7313 \\
\hline \hline
\multicolumn{7}{l}{\footnotesize{Note: The DNN-CS\textsubscript{VT} model uses the vanilla topk (VT) function for CS.}} \\
\end{tabular}
\end{table}

In contrast to the DNN and DNN-CS\textsubscript{VT} models, the use of the DNN-CS network enabled ElectrodeNet-CS to achieve success in both compatibility and performance with the ACE strategy.
In the DNN-CS model, the custom topk layer not only learned to minimize the loss with the M-channel envelopes by selecting N\textsubscript{topk} maxima and suppressing the other positive network outputs, but also adjusted the remaining negative outputs with high loss to be non-negative during training.
As a result, ElectrodeNet-CS increased the weights important to speech features, weakened connections less relevant to the loss function, preserved the N-of-M compatible pattern without bypassing any channels, and produced the specified number of maxima (N\textsubscript{CS} $\leq$ N\textsubscript{topk}).
With the special CS design, both the objective and subjective evaluation results showed that the ElectrodeNet-CS strategy had similar or even slightly higher average scores than ACE in some conditions.
Specifically, the higher mean STOI scores across all noisy conditions and the higher mean sentence recognition results for at -5 dB SNR are worthy to be noticed.
The inspiring findings in ElectrodeNet-CS encouraged further exploration of the mechanisms of deep learning based envelope detection and channel selection.
Furthermore, the flexibility of network architecture to incorporate new functions is an important strength of deep learning for future CI research.

\subsection{General Discussion}

The utilization of deep learning in CI coding strategy has provided new insights into the ‘AI + CI’ field.
AI not only transforms hearing healthcare and research \cite{AI_hearing_review_lesica2021}, but also has great potential to break through the present limitations with CI coding strategy.
With more advanced network architectures and training methods, deep learning based coding strategies may outperform the present coding strategies.
Moreover, deep learning can be considered not as a competitor, but as a collaborator in complement to traditional signal processing approaches.
Researchers may use deep learning not only as a tool to resolve problems, but also as an approach to explore undiscovered knowledge.
Based on the present findings contributed by numerous experts, it is promising to see the potential for deep learning to benefit CI users in the future.

On the journey toward a deep learning based sound coding strategy, issues of complexity, latency, safety, and ethics are important for researchers.
With the advances in calculation speed and energy efficiency, computational complexity is becoming less of a primary issue in hardware design, especially for compressed pre-trained network models on devices.
Using the analogy of commercial hearing aids with DNN models \cite{Oticon_beck2021} and complicated ASR chips \cite{ASR_IC_litovsky2017}, the innovation for CI coding strategy is expected to be in a similar transitional process.
For hearing assistive devices, latency is another critical challenge, and real-time inference with compact models is an important direction for investigation \cite{DL_edge_chen2019}.
As a neural prosthetic device is implanted in the human body, thorough considerations of safety and ethics are essential for deep learning driven algorithms.
For example, the maxima selection, presentation level, and pulse patterns determined using deep learning must comply with the safety limits to prevent biological damage and uncomfortable over-stimulations \cite{CI_tool_review_litovsky2017}.
In addition, AI raises ethical issues, such as privacy concerns in collecting user data for re-training models and further research.
Equal and effective communication is required between researchers, practitioners, and CI users in collecting personal information and expectations in AI technologies.
Present guidelines related to research ethics and IRB approvals may need to be renovated to reinforce moral values, considerations, and regulations for researchers and institutes in the development of deep learning based neural prosthetic devices \cite{AI_ethics_hagendorff2020, audiology_ethics_wasmann2021}.

More investigations on deep learning based coding strategies are necessary.
The present study focused on mimicking the core processing of the ACE strategy using neural networks and hence did not consider using noisy, reverberant, and mixed speech as input data to train the network model.
It is planned to build a more powerful neural-based CI system, which can flexibly combine ElectrodeNet with various speech preprocessing, such as speech denoising, dereverberation, and speech separation, using more advanced network architectures and training methods, such as end-to-end learning and joint-training.
Additional multimodal information, such as contralateral sounds, visual cues, tactile vibrations, and auditory and vestibular responses, may also be processed in the same system \cite{CROS_kurien2019, ETS_zeng2017, citi_visualcues_ci2021, CVI_Lanthaler2021}.
Therefore, expanding deep learning modules is potential to provide breakthroughs in CI signal processing and there is a great demand for further research by conducting more simulations and clinical experiments with CI listeners.
The innovation for ElectrodeNet is on-going to make some differences to the ‘AI + CI’ and other relevant fields.

\section{Conclusion}

Investigations on ElectrodeNet revealed that deep learning based sound coding strategies may have similar or even slightly higher average speech intelligibility compared to the ACE strategy.
By replacing the conventional envelope detection and CS stages using a neural network, the ElectrodeNet and ElectrodeNet-CS strategies performed satisfactorily:

\begin{enumerate}
\item Strong correlations exhibited between ACE and the DNN, CNN, and LSTM based ElectrodeNets based on the MSE, LCC, and SRCC.
\item Strong correlations exhibited between ACE and the DNN based ElectrodeNet for both TMHINT and TIMIT datasets, and for the four types of noises.
\item The ElectrodeNet-CS strategy was capable of producing N-of-M compatible electrode stimulation patterns.
\item After combining the CS function, ElectrodeNet-CS performed similarly or slightly above average in STOI and sentence recognition results compared to ACE.
\end{enumerate}

The findings suggest the substantial feasibility of using ElectrodeNet as an alternative to traditional coding strategies and its flexible network architecture to create new networks, such as ElectrodeNet-CS, to achieve higher performance.
This study provides some insights and research directions toward the inevitable trend of the `AI + CI’ revolution.

\section{Appendix}

\begin{figure*}[]
\centering{\includegraphics{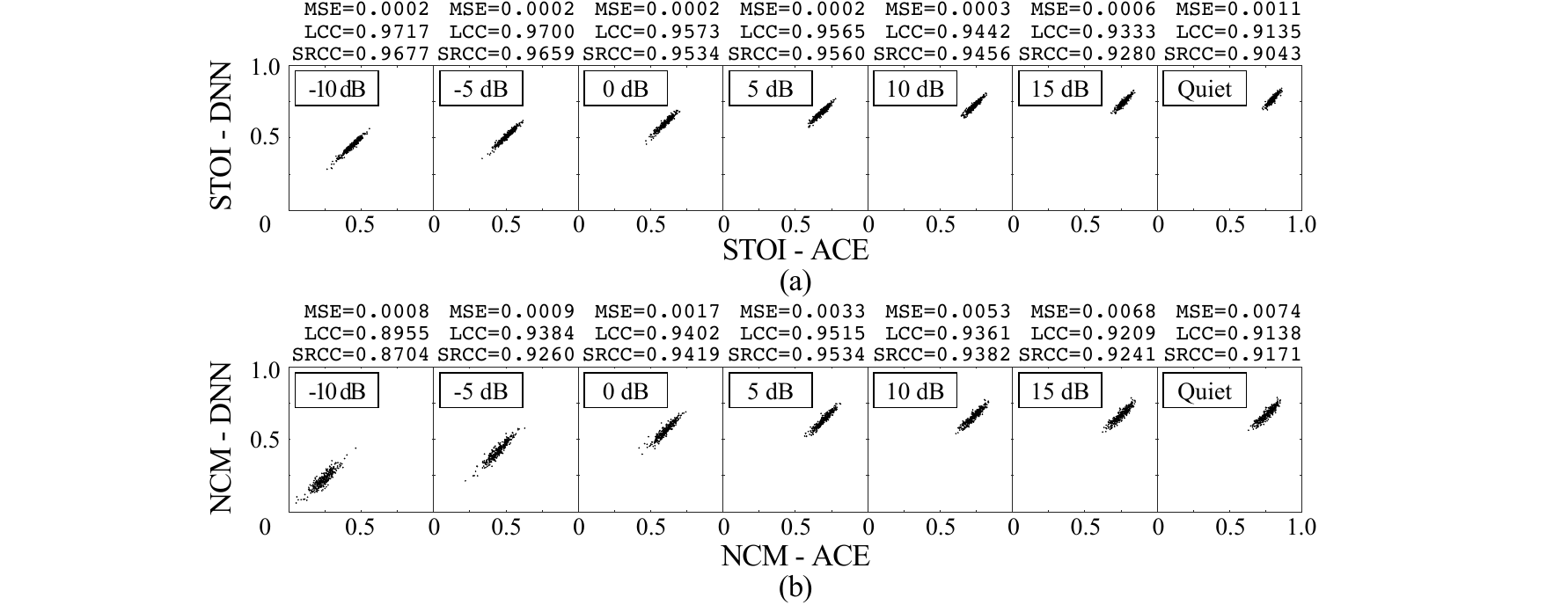}}
\caption{Scatter plots of (a) STOI and (b) NCM scores at different SNR levels for the ACE strategy and the DNN based ElectrodeNet strategy with N = 12.}
\label{fig7_Appendix_Scatter_bySNR}
\end{figure*}

To demonstrate the correlations between the ACE strategy and the DNN based ElectrodeNet in STOI and NCM scores for each SNR level, Fig.\,\ref{fig7_Appendix_Scatter_bySNR} shows their scatter plots and correlation scores.
All dots in each sub-figure are arranged close to a positive linear relationship.
The correlation scores indicate strong correlations between the two strategies (MSEs $\leq$ 0.0011, LCCs $\geq$ 0.9135, SRCCs $\geq$ 0.9043 for STOI, and MSEs $\leq$ 0.0074, LCCs $\geq$ 0.8955, SRCCs $\geq$ 0.8704 for NCM).
The MSEs are slightly larger than those computed across all SNR levels in Fig.\,\ref{fig3_ScatterSTOI_NCM} (a) and (d) (MSEs = 0.0004 for STOI and MSE = 0.0041 for NCM), where correlation scores may be driven both by the characters of neural networks and the SNR levels. 
The correlation coefficients for each SNR level are slightly smaller than those in Fig.\,\ref{fig3_ScatterSTOI_NCM} (a) and (d) (LCC = 0.9949 and SRCC = 0.9921 for STOI, and LCC = 0.9919, SRCC = 0.9807 for NCM).
Although the scores computed across different SNR level slightly enhance correlation scores, Fig.\,\ref{fig7_Appendix_Scatter_bySNR} shows that strong correlations between ACE and ElectrodeNet are still present at single SNR levels.

\balance


\begin{thebibliography}{1}

\bibitem{CI_review_zeng2008}
F.-G. Zeng, S. Rebscher, W. Harrison, X. Sun, and H. Feng, ``Cochlear implants: system design, integration, and evaluation,” \emph{IEEE Rev. Biomed. Eng.}, vol. 1, pp. 115–142, 2008.

\bibitem{CI_review_clark2015}
G. M. Clark, ``The multi-channel cochlear implant: Multi-disciplinary development of electrical stimulation of the cochlea and the resulting clinical benefit,” \emph{Hear. Res.}, vol. 322, pp. 4–13, 2015.

\bibitem{CI_review_wilson2015}
B. S. Wilson, ``Getting a decent (but sparse) signal to the brain for users of cochlear implants,” \emph{Hear. Res.}, vol. 322, pp. 24–38, 2015.

\bibitem{CI_challenges_TBME_zeng2017}
F.-G. Zeng, ``Challenges in improving cochlear implant performance and accessibility,” \emph{IEEE Trans. Biomed. Eng.}, vol. 64, no. 8, pp. 1662–1664, 2017.

\bibitem{CI_cognition_adult_moberly2019}
A. C. Moberly, K. Doerfer, and M. S. Harris, ``Does cochlear implantation improve cognitive function?" \emph{Laryngoscope}, vol. 129, no. 10, pp. 2208–2209, 2019

\bibitem{CI_cognition_children_almomani2021}
F. Almomani \emph{et al.}, ``Cognitive functioning in deaf children using cochlear implants," \emph{BMC pediatrics}, vol. 21, no. 1, pp. 1-13, 2021.

\bibitem{Covid_children_HL_Taddei2021}
A. Taddei, E. A. López, and R. A. R. Reyes, ``Children with hearing disabilities during the pandemic: Challenges and perspectives of inclusion.'' \emph{Educ. Sci. Society-Open Access}. vol. 12, no. 1, pp.178–196, 2021.

\bibitem{Covid_adult_CI_Perez2022}
F. P. P{\'e}rez, D. E. H. Hartley, P. Kitterick, and I. M. Wiggins, ``Perceived listening difficulties of adult cochlear-implant users under measures introduced to combat the spread of COVID-19.'' \emph{Trends Hear.}, vol. 26, pp.1–22, 2022.

\bibitem{DL_review_lecun2015}
Y. LeCun, Y. Bengio, and G. Hinton, ``Deep learning,” \emph{Nature}, vol. 521, no. 7553, pp. 436–444, 2015.

\bibitem{DL_book_goodfellow2016}
I. Goodfellow, Y. Bengio, and A. Courville, \emph{Deep learning}. MIT press, 2016.

\bibitem{AI_hearing_review_lesica2021}
N. A. Lesica, N. Mehta, J. G. Manjaly, L. Deng, B. S. Wilson, and F.-G. Zeng, ``Harnessing the power of artificial intelligence to transform hearing healthcare and research,” \emph{Nat. Mach. Intell.}, vol. 3, no. 10, pp. 840–849, 2021.

\bibitem{ML_review_crowson2020}
M. G. Crowson, V. Lin, J. M. Chen, and T. C. Chan, ``Machine learning and cochlear implantation—A structured review of opportunities and challenges,” \emph{Otol. Neurotol.}, vol. 41, no. 1, pp. e36–e45, 2020.

\bibitem{CI_DataMining_prognosis_Guerra_Jimenez2016}
G. Guerra-Jim{\'e}nez, {\'A}. R. De Miguel, J. C. F. Gonz{\'a}lez, S. A. B. Barreiro, D. P. Plasencia, and {\'A}. R. Mac{\'\i}as, ``Cochlear implant evaluation: Prognosis estimation by data mining system," \emph{J. Int. Adv. Otol.}, vol. 12, no. 1, pp. 1-7, 2016.

\bibitem{CI_AI_TBME_electrode_gao2016}
X. Gao, D. B. Grayden, and M. D. McDonnell, ``Modeling electrode place discrimination in cochlear implant stimulation,” \emph{IEEE Trans. Biomed. Eng.}, vol. 64, no. 9, pp. 2219–2229, 2016.

\bibitem{ML_SVM_Surgery_pile2017}
J. Pile, G. B. Wanna, and N. Simaan, ``Robot-assisted perception augmentation for online detection of insertion failure during cochlear implant surgery,” \emph{Robotica}, vol. 35, no. 7, pp. 1598–1615, 2017.

\bibitem{ML_mapping_meeuws2017}
M. Meeuws, D. Pascoal, I. Bermejo, M. Artaso, G. De Ceulaer, and P. J. Govaerts, ``Computer-assisted {CI} fitting: Is the learning capacity of the intelligent agent fox beneficial for speech understanding?” \emph{Cochlear Implants Int.}, vol. 18, no. 4, pp. 198–206, 2017.

\bibitem{ML_detect_reverb_desmond2013}
J. M. Desmond, L. M. Collins, and C. S. Throckmorton, ``Using channel-specific statistical models to detect reverberation in cochlear implant stimuli,” \emph{J. Acoust. Soc. Amer.}, vol. 134, no. 2, pp. 1112–1120, 2013.

\bibitem{ML_dereverb_chu2018}
K. Chu, C. Throckmorton, L. Collins, and B. Mainsah, ``Using machine learning to mitigate the effects of reverberation and noise in cochlear implants,” in \emph{Proc. Mtgs. Acoust.}, vol. 33, no. 1. pp. 1–13. 2018,

\bibitem{ML_acoustics_bianco2019}
M. J. Bianco \emph{et al.}, ``Machine learning in acoustics: Theory and applications,” \emph{J. Acoust. Soc. Amer.}, vol. 146, no. 5, pp. 3590–3628, 2019.

\bibitem{DL_audio_review_purwins2019}
H. Purwins, B. Li, T. Virtanen, J. Schlüter, S.-Y. Chang, and T. Sainath, ``Deep learning for audio signal processing,” \emph{IEEE J. Sel. Top. Signal. Process.}, vol. 13, no. 2, pp. 206–219, 2019.

\bibitem{DL_HA_wang2017}
D. Wang, ``Deep learning reinvents the hearing aid," \emph{IEEE Spectr.}, vol. 54, no. 3, pp. 32-37, 2017.

\bibitem{Oticon_beck2021}
D. Beck, ``Hearing, listening and deep neural networks in hearing aids,” \emph{J. Otolaryngol.-ENT Res.}, vol. 13, no. 1, pp. 5–8, 2021.

\bibitem{NNSE_bolner2016}
F. Bolner, T. Goehring, J. Monaghan, B. Van Dijk, J. Wouters, and S. Bleeck, ``Speech enhancement based on neural networks applied to cochlear implant coding strategies,” in \emph{Proc. IEEE Int. Conf. Acoust. Speech Signal Process. (ICASSP)}. 2016, pp. 6520–6524.

\bibitem{NNSE_goehring2017}
T. Goehring, F. Bolner, J. J. Monaghan, B. Van Dijk, A. Zarowski, and S. Bleeck, ``Speech enhancement based on neural networks improves speech intelligibility in noise for cochlear implant users,” \emph{Hear. Res.}, vol. 344, pp. 183–194, 2017.

\bibitem{citi_ci2017}
Y.-H. Lai, F. Chen, S.-S. Wang, X. Lu, Y. Tsao, and C.-H. Lee, ``A deep denoising autoencoder approach to improving the intelligibility of vocoded speech in cochlear implant simulation,” \emph{IEEE Trans. Biomed. Eng.}, vol. 64, no. 7, pp. 1568–1578, 2017.

\bibitem{citi_ci2018}
Y.-H. Lai \emph{et al.}, ``Deep learning–based noise reduction approach to improve speech intelligibility for cochlear implant recipients,” \emph{Ear Hear.}, vol. 39, no. 4, pp. 795–809, 2018.

\bibitem{CNN_Interspeech_mamun2019}
N. Mamun, S. Khorram, and J. H. Hansen, ``Convolutional neural network-based speech enhancement for cochlear implant recipients,” in \emph{Proc. Interspeech}, 2019, pp. 4265–4269.

\bibitem{citi_eas_ci2021}
N. Y.-H. Wang \emph{et al.}, ``Improving the intelligibility of speech for simulated electric and acoustic stimulation using fully convolutional neural networks,” \emph{IEEE Trans. Neural Syst. Rehabil. Eng.}, vol. 29, pp. 184–195, 2021.

\bibitem{citi_visualcues_ci2021}
R.-Y. Tseng, T.-W. Wang, S.-W. Fu, C.-Y. Lee, and Y. Tsao, ``A study of joint effect on denoising techniques and visual cues to improve speech intelligibility in cochlear implant simulation,” \emph{IEEE Trans. Cogn. Develop. Syst.}, vol. 13, no. 4, pp. 984-994, 2021.

\bibitem{DRNN_nogueira2016}
W. Nogueira, T. Gajecki, B. Krüger, J. Janer, and A. Büchner, ``Development of a sound coding strategy based on a deep recurrent neural network for monaural source separation in cochlear implants,” in \emph{Proc. of the 12th ITG Symp. on Speech Comm.}, 2016.

\bibitem{SE_RNN_goehring2019}
T. Goehring, M. Keshavarzi, R. P. Carlyon, and B. C. Moore, ``Using recurrent neural networks to improve the perception of speech in non- stationary noise by people with cochlear implants,” \emph{J. Acoust. Soc. Amer.}, vol. 146, no. 1, pp. 705–718, 2019.

\bibitem{LSTM_CI_2021}
K. Chu, L. Collins, and B. Mainsah, ``A causal deep learning framework for classifying phonemes in cochlear implants," In \emph{Proc. IEEE Int. Conf. Acoust. Speech Signal Process. (ICASSP)}, 2021, pp. 6498-6502.

\bibitem{remix_nogueira2016}
J. Pons, J. Janer, T. Rode, and W. Nogueira, ``Remixing music using source separation algorithms to improve the musical experience of cochlear implant users,” \emph{J. Acoust. Soc. Amer.}, vol. 140, no. 6, pp. 4338–4349, 2016.

\bibitem{remix_nogueira2018}
T. Gajecki and W. Nogueira, ``Deep learning models to remix music for cochlear implant users,” \emph{J. Acoust. Soc. Amer.}, vol. 143, no. 6, pp. 3602–3615, 2018.

\bibitem{remix_review_nogueira2018}
W. Nogueira, A. Nagathil, and R. Martin, ``Making music more accessible for cochlear implant listeners: Recent developments,” \emph{IEEE Signal Process. Mag.}, vol. 36, no. 1, pp. 115–127, 2018.

\bibitem{strategy_review_wouters2015}
J. Wouters, H. J. McDermott, and T. Francart, ``Sound coding in cochlear implants: From electric pulses to hearing,” \emph{IEEE Signal Process. Mag.}, vol. 32, no. 2, pp. 67–80, 2015.

\bibitem{CIS_wilson1991}
B. S. Wilson, C. C. Finley, D. T. Lawson, R. D. Wolford, D. K. Eddington, and W. M. Rabinowitz, ``Better speech recognition with cochlear implants,” \emph{Nature}, vol. 352, no. 6332, pp. 236–238, 1991.

\bibitem{ACE_original_n_of_m_vandali2000}
A. E. Vandali, L. A. Whitford, K. L. Plant, G. M. Clark, ``Speech perception as a function of electrical stimulation rate: Using the Nucleus 24 cochlear implant system,” \emph{Ear Hear.}, vol. 21, no. 6, pp. 608–624, 2000.

\bibitem{MP3000_PACE_nogueira2005}
W. Nogueira, A. B{\"u}chner, T. Lenarz, and B. Edler, ``A psychoacoustic ``NofM”-type speech coding strategy for cochlear implants,” \emph{EURASIP J. Adv. Signal Process.}, vol. 2005, no. 18, pp. 3044–3095, 2005.

\bibitem{FSP_riss2011}
D. Riss \emph{et al.}, ``Envelope versus fine structure speech coding strategy: A crossover study,” \emph{Otol. Neurotol.}, vol. 32, no. 7, pp. 1094–1101, 2011.

\bibitem{HiRes120_brendel2008}
M. Brendel, A. Buechner, B. Krueger, C. Frohne-Buechner, and T. Lenarz, ``Evaluation of the Harmony sound processor in combination with the speech coding strategy HiRes 120,” \emph{Otol. Neurotol.}, vol. 29, no. 2, pp. 199–202, 2008.

\bibitem{HiRes120_vs_HiRes_firszt2009}
J. B. Firszt, L. K. Holden, R. M. Reeder, and M. W. Skinner, ``Speech recognition in cochlear implant recipients: Comparison of standard HiRes and HiRes 120 sound processing,” \emph{Otol. Neurotol.}, vol. 30, no. 2, pp. 146–152, 2009.

\bibitem{OticonNeuro_CRYSTALIS_schramm2020}
D. Schramm \emph{et al.}, ``Clinical efficiency and safety of the Oticon Medical Neuro cochlear implant system: A multicenter prospective longitudinal study,” \emph{Expert Rev. Med. Devices}, vol. 17, no. 9, pp. 959–967, 2020.

\bibitem{NCU_CI_TNSRE_huang2021}
E. H.-H. Huang, C.-M. Wu, and H.-C. Lin, ``Combination and comparison of sound coding strategies using cochlear implant simulation with Mandarin speech," \emph{IEEE Trans. Neural Syst. Rehabil. Eng.}, vol. 29, pp. 2407–2416, 2021.

\bibitem{single_ch_SE_hansne2010}
S. O. Sadjadi and J. H. Hansen, ``Assessment of single-channel speech enhancement techniques for speaker identification under mismatched conditions,” in \emph{Proc. Interspeech}, 2010, pp. 2138–2142.

\bibitem{one_pass_fujimoto2019}
M. Fujimoto and H. Kawai, ``One-pass single-channel noisy speech recognition using a combination of noisy and enhanced features,” in \emph{Proc. Interspeech}, 2019, pp. 486–490.

\bibitem{input_switch_sato2022}
H. Sato, T. Ochiai, M. Delcroix, K. Kinoshita, N. Kamo, and T. Moriya, ``Learning to enhance or not: Neural network-based switching of enhanced and observed signals for overlapping speech recognition," In \emph{Proc. IEEE Int. Conf. Acoust. Speech Signal Process. (ICASSP)}, 2022, pp. 6287-6291.

\bibitem{NCU_CI_ElectrodeNet_huang2019}
E. H.-H. Huang, K.-H. Hung, Y. Tsao, and C.-M. Wu, ``ElectrodeNet – artificial intelligence based sound coding strategy for cochlear implants,” in \emph{Proc. of the 12th Asia Pacific Symp. Cochlear Implants Rel. Sci. (APSCI2019)}. Tokyo, Japan, 2019, p. O2-5.

\bibitem{DL_edge_chen2019}
J. Chen and X. Ran, ``Deep learning with edge computing: A review,” \emph{Proc. IEEE}, vol. 107, no. 8, pp. 1655—1674, Aug. 2019.

\bibitem{PhDthesis_pitch_swanson2008}
B. A. Swanson, ``Pitch perception with cochlear implants,” Ph.D. dissertation, Dept. Otol., Univ. Melbourne, VIC, Australia, 2008.

\bibitem{vocoder_shannon1995}
R. V. Shannon, F.-G. Zeng, V. Kamath, J. Wygonski, and M. Ekelid, ``Speech recognition with primarily temporal cues,” \emph{Science}, vol. 270, no. 5234, pp. 303–304, 1995.

\bibitem{vocoder_tone_dorman2002}
M. F. Dorman, P. C. Loizou, A. J. Spahr, and E. Maloff, ``A comparison of the speech understanding provided by acoustic models of fixed-channel and channel-picking signal processors for cochlear implants,” \emph{J. Speech Lang. Hear. Res.}, vol. 44, pp. 1–6, 2002.

\bibitem{NMT_swanson2006}
B. Swanson and H. Mauch, \emph{Nucleus Matlab Toolbox 4.20 software user manual}, Cochlear Ltd, Lane Cove NSW, Australia, 2006.

\bibitem{DNN_BP_werbos1981}
P. J. Werbos, ``Applications of advances in nonlinear sensitivity analysis," in \emph{Proc. of the 10th IFIP Conf.}, New York City, pp. 762-770, 1981.

\bibitem{MLP_Hornik1989}
K. Hornik, M. Stinchcombe, and H. White, ``Multilayer feedforward networks are universal approximators," \emph{Neural Netw.}, vol. 2, no.5, pp. 359-366, 1989.

\bibitem{CNN_lecun1989}
Y. LeCun, ``Generalization and network design strategies," Technical Report CRG-TR-89-4, Univ. Toronto, Canada, pp. 1–19, 1989.

\bibitem{LSTM_hochreiter1997}
S. Hochreiter and J. Schmidhuber, ``Long short-term memory," \emph{Neural Comput.}, vol. 9, no. 8, pp. 1735–1780, 1997.

\bibitem{RNN_rumelhart1986}
D. E. Rumelhart, G. E. Hinton, and R. J. Williams, ``Learning representations by back-propagating errors," \emph{Nature}, vol. 323, no. 9, 533-536, 1986.

\bibitem{ReLU_maas2013}
A. L. Maas, A. Y. Hannun, and A. Y. Ng, ``Rectifier nonlinearities improve neural network acoustic models," in \emph{Proc. ICML}, 2013, pp. 1–6.

\bibitem{Adam_kingma2015}
D. P. Kingma, and J. Ba. ``Adam: A method for stochastic optimization," in \emph{Proc. ICLR}, May 2015.

\bibitem{PyTorch_topk_web}
PyTorch, ``TORCH.TOPK,” Accessed: Nov. 22, 2022. [Online] Available: https://pytorch.org/docs/stable/generated/torch.topk.html

\bibitem{MHINT_wong2007}
L. L. Wong, S. D. Soli, S. Liu, N. Han, and M.-W. Huang, ``Development of the Mandarin hearing in noise test (MHINT),” \emph{Ear Hear.}, vol. 28, no. 2, pp. 70S–74S, 2007.

\bibitem{TIMIT_garofolo1993}
J. S. Garofolo, L. F. Lamel, W. M. Fisher, J. G. Fiscus, and D. S. Pallett, ``DARPA TIMIT acoustic-phonetic continuous speech corpus CD-ROM," \emph{NASA STI/Recon Tech. Rep. N}, vol. 93, no. 27403, 1993.

\bibitem{Mandarin_trisyllabic_word_nissen2007}
S. L. Nissen, R. W. Harris, and K. B. Slade, ``Development of speech reception threshold materials for speakers of Taiwan Mandarin,” \emph{Int. J. Audiol.}, vol. 46, no. 8, pp. 449–458, 2007.

\bibitem{stoi_taal2011}
C. H. Taal, R. C. Hendriks, R. Heusdens, and J. Jensen, ``An algorithm for intelligibility prediction of time-frequency weighted noisy speech,” \emph{IEEE/ACM Trans. Audio Speech Lang. Process.}, vol. 19, no. 7, pp. 2125–2136, 2011.

\bibitem{ncm_holube1996}
I. Holube and B. Kollmeier, ``Speech intelligibility prediction in hearing-impaired listeners based on a psychoacoustically motivated perception model,” \emph{J. Acoust. Soc. Amer.}, vol. 100, no. 3, pp. 1703–1716, 1996.

\bibitem{ncm_goldsworthy2004}
R. L. Goldsworthy and J. E. Greenberg, ``Analysis of speech-based speech transmission index methods with implications for nonlinear operations,” \emph{J. Acoust. Soc. Amer.}, vol. 116, no. 6, pp. 3679–3689, 2004.

\bibitem{ncm_evaluate_chen2011}
F. Chen and P. C. Loizou, ``Predicting the intelligibility of vocoded and wideband Mandarin Chinese,” \emph{J. Acoust. Soc. Amer.}, vol. 129, no. 5, pp. 3281–3290, 2011.

\bibitem{SRMR_objective_review_loizou_santos2013}
J. F. Santos, S. Cosentino, O. Hazrati, P. C. Loizou, and T. H. Falk,
``Objective speech intelligibility measurement for cochlear implant users in complex listening environments,” \emph{Speech Commun.}, vol. 55, no. 7-8, pp. 815–824, 2013.

\bibitem{objective_evaluation_review_falk2015}
T. H. Falk \emph{et al.}, ``Objective quality and intelligibility prediction for users of assistive listening devices: Advantages and limitations of existing tools,” \emph{IEEE Signal Process. Mag.}, vol. 32, no. 2, pp. 114–124, 2015.

\bibitem{OSNR_watkins2018}
G. D. Watkins, B. A. Swanson, and G. J. Suaning, ``An evaluation of output signal to noise ratio as a predictor of cochlear implant speech intelligibility,” \emph{Ear Hear.}, vol. 39, no. 5, pp. 958–968, 2018.

\bibitem{Sparse_NMF_hu2015}
H. Hu, M. E. Lutman, S. D. Ewert, G. Li, and S. Bleeck, ``Sparse nonnegative matrix factorization strategy for cochlear implants,” \emph{Trends Hear.}, vol. 19, pp. 1–16, 2015.

\bibitem{MOCstrategy_lopez_poveda2018}
E. A. Lopez-Poveda, and A. Eustaquio-Mart{\'\i}n, “Objective speech transmission improvements with a binaural cochlear implant sound-coding strategy inspired by the contralateral medial olivocochlear reflex,” \emph{J. Acoust. Soc. Amer.}, vol. 143, no. 4, pp. 2217–2231, 2018.

\bibitem{MOCstrategy_lopez_poveda2020}
E. A. Lopez-Poveda \emph{et al.}, ``Speech-in-noise recognition with more realistic implementations of a binaural cochlear-implant sound coding strategy inspired by the medial olivocochlear reflex,” \emph{Ear Hear.}, vol. 41, no. 6, pp. 1492–1510, 2020.

\bibitem{VG_langner2020}
F. Langner, A. B{\"u}chner, and W. Nogueira, ``Evaluation of an adaptive dynamic compensation system in cochlear implant listeners,” \emph{Trends Hear.}, vol. 24, pp. 1–13, 2020.

\bibitem{TFMask_mourao2020}
G. L. Mour{\~a}o, M. H. Costa, and S. Paul, ``Speech intelligibility for cochlear implant users with the MMSE noise-reduction time-frequency mask,” \emph{Biomed Signal Process Control}, vol. 60, p. 101982, 2020.

\bibitem{SE_book_loizou2013}
P. C. Loizou, \emph{Speech Enhancement: Theory and Practice}, 2nd ed., Boca Raton, FL, USA: CRC Press, 2013.

\bibitem{LCC_pearson1920}
K. Pearson, ``Notes on the history of correlation,” \emph{Biometrika}, vol. 13, no. 1, pp. 25–45, 1920.

\bibitem{SRCC_spearman1904}
C. Spearman, ``The proof and measurement of association between two things,” \emph{Am. J. Psychol.}, vol. 15, no. 1, pp. 72–101, 1904.

\bibitem{Correlation_Strong_0p8_zou2003}
K. H. Zou, K. Tuncali, and S. G. Silverman, ``Correlation and simple linear regression." \emph{Radiology}, vol. 227, no. 3, pp. 617–628, 2003.

\bibitem{Correlation_Strong_0p7_akoglu2018}
H. Akoglu, ``User's guide to correlation coefficients." \emph{Turkish J. Emerg. Med.}, vol. 18, no. 3, pp. 91–93, 2018.

\bibitem{NCU_CI_wu2009}
C.-M. Wu, K.-Y. Huang, and H.-C. Lin, ``Effects of channel number, stimulation rate, and electroacoustic stimulation of cochlear implant simulation on Chinese speech recognition in noise,” in \emph{Proc. of the 7-th Asia Pacific Symp. Cochlear Implants Rel. Sci. (APSCI2009)}, Singapore, 2009, p. RS2B-7.

\bibitem{ASR_IC_litovsky2017}
M. Price, J. Glass, and A. P. Chandrakasan, ``A scalable speech recognizer with deep-neural-network acoustic models and voice-activated power gating," In \emph{2017 Proc. IEEE Int. Solid-State Circuits Conf. (ISSCC)}, pp. 244-245, 2017.

\bibitem{CI_tool_review_litovsky2017}
R. Y. Litovsky, M. J. Goupell, A. Kan, and D. M. Landsberger, ``Use of research interfaces for psychophysical studies with cochlear-implant users," \emph{Trends Hear.}, vol. 21, pp. 1–15, 2017.

\bibitem{AI_ethics_hagendorff2020}
T. Hagendorff, ``The ethics of AI ethics: An evaluation of guidelines," \emph{Minds Mach.}, vol. 30, no. 1, pp. 99–120, 2020.

\bibitem{audiology_ethics_wasmann2021}
J. W. Wasmann \emph{et al.}, ``Computational audiology: New approaches to advance hearing health care in the digital age," \emph{Ear Hear}, vol. 42, no. 6, pp. 1499–1507, 2021.

\bibitem{CROS_kurien2019}
G. Kurien \emph{et al.}, ``The benefit of a wireless contralateral routing of signals (CROS) microphone in unilateral cochlear implant recipients," \emph{Otol. Neurotol.}, vol. 40, no. 2, e82-e88, 2019.

\bibitem{ETS_zeng2017}
J. Huang, B. Sheffield, P. Lin, and F. G. Zeng, ``Electro-tactile stimulation enhances cochlear implant speech recognition in noise," \emph{Sci. Rep.}, vol. 7, no. 1, pp. 1–5, 2017.

\bibitem{CVI_Lanthaler2021}
D. Lanthaler \emph{et al.}, ``Speech perception with novel stimulation strategies for combined cochleo-vestibular systems," \emph{IEEE Trans. Neural Syst. Rehabil. Eng.}, vol. 29, pp. 1644–1650, 2021.

\end{thebibliography}
\end{document}